\documentclass[12pt,preprint]{aastex}
\usepackage{epsf,epsfig}

\def\etal{{\it et al.\ }}
\def\iso#1#2{\mbox{${}^{#2}{\rm #1}$}}
\newcommand\he[1]{\iso{He}{#1}}
\newcommand\li[1]{\iso{Li}{#1}}
\newcommand\hii{\ion{H}{2}}
\def\c1#1{\iso{C}{1#1}}

\begin{document}

\title {On the baryometric status of $^{3}{\rm He}$}

\author{Elisabeth Vangioni-Flam}
\affil{Institut d'Astrophysique, 98 bis Boulevard Arago,
Paris 75014, France}

\author{Keith~A.~Olive}
\affil{ Theoretical Physics
Institute, School of Physics and Astronomy, \\
University of Minnesota, Minneapolis, MN 55455 USA}

\author{Brian D. Fields}
\affil{Department of Astronomy, University of Illinois,
Urbana, IL 61801, USA}

\and \author{Michel Cass\'e}
\affil{Service d'Astrophysique, CEA, Orme des Merisiers,
91191 Gif sur Yvette, France \\
also Institut d'Astrophysique, 98 bis Boulevard Arago,
Paris 75014, France}

\begin{abstract}

\vskip-5.5in
\begin{flushright}
UMN-TH-2106/02 \\
TPI-MINN-02/21 \\
astro-ph/0207583 \\
July 2002
\end{flushright}
\vskip+4.4in

Recent observations by Bania \etal (2002) 
measure \he3 versus oxygen in Galactic \hii\ regions,
finding that \he3/H is within a factor of 2 of the
solar abundance for [O/H] $\ga -0.6$.
These results are consistent with a flat
behavior in this metallicity range,
tempting one to deduce from these observations a primordial
value for the \he3 abundance,
which could join D and \li7 as an indicator of
the cosmic baryon density.
However, using the same data, we show that it is not
possible to obtain a strong constraint on the baryon
density range. This is due to (i) the intrinsically weak sensitivity of 
the primordial \he3 abundance to the baryon
density; (ii) the limited range in metallicity of the
sample; (iii) the intrinsic scatter in the data; and  (iv) our limited
understanding of the chemical and stellar evolution of this isotope. 
Consequently, the
\he3 observations correspond to an extended range of baryon-to-photon
ratio,
$\eta = (2.2 - 6.5)\times 10^{-10}$, which diminishes the role of
\he3 as a precision baryometer. On the other hand, 
once the baryon-to-photon ratio is determined by the CMB, D/H, or \li7/H,
the primordial value of \he3/H can be inferred.
Henceforth new observations of Galactic \he3,
can in principle greatly improve our understanding of
stellar and/or chemical evolution and reconcile the observations of
the \hii\ regions and those of the planetary nebulae.

\end{abstract}


\section{Introduction}

As the sole parameter of the underlying theory of big bang
nucleosynthesis (BBN), the baryon-to-photon density ratio,
$\eta \propto \Omega_{\rm B} h^2$, is one of the holy grails of cosmology.  In the past,
$\eta$ was best determined by the concordance of the four light
element isotopes produced by BBN, D, \he3, \he4, and \li7
(Walker \etal 1991; Schramm \& Turner 1998; Olive, Steigman, \&
Walker 2000; 
Nollett \& Burles 2000;
Cyburt, Fields, \& Olive 2001; Coc \etal 2002; 
Fields \& Sarkar 2002).  As the systematic uncertainties in the abundance determination
of each of these isotopes are becoming better understood, it seems that our
ability to `predict' a precise value of $\eta$ diminishes. Perhaps, our best
hope for an accurate determination of $\eta$ lies with the analysis of
the microwave background anisotropy power spectrum. From this independent
determination, we can certainly expect to gain a substantial amount of
insight in the systematic effects involved in the abundance measurements
(Kneller \etal 2001; Cyburt, Fields, \& Olive 2002). Of course, the
concordance of BBN (within known uncertainties) remains a critical test of
the standard cosmological model up to temperature scales of order $\ga 1$
MeV. Needless to say, BBN continues to provide countless constraints on
particle physics models which affect the evolution of the Universe at
that epoch. 

It is in this context that Bania \etal (2002) have 
recently measured \he3/H in about
20 \hii\ regions in the Galactic disk.  
These data show almost no variation over the metallicity range [O/H] =
$-0.6$ to +0.2, i.e. a `plateau' in [\he3/H]
vs  [O/H].
Using recent developments of
the stellar evolution theory, these authors have set an upper limit to the
primordial \he3 abundance 
of ${\rm \he3/H} \le (1.1 \pm 0.2) \times 10^{-5}$, corresponding to $\Omega_{b}h^{2}$ of about
0.02.  Bania \etal\ argue that the upper limit is robust despite the
uncertainties in the details of \he3 evolution; this
robustness is argued to restore \he3 as a BBN baryometer.

However, in the nineties severe doubt was cast regarding the use of \he3
as a baryon density indicator due to the large uncertainties in its
production in low mass stars. Standard stellar theory for low mass stars
(see e.g. Iben \& Truran, 1978) predicts a significant amount of
production in these stars.  When incorporated into simple models of
Galactic chemical evolution, one would expect \he3 abundances in great
excess from those observed (Vangioni-Flam \etal 1994; Olive \etal 1995;
Galli
\etal 1995; Scully \etal 1996, 1997; Dearborn, Steigman, \& Tosi
1996). While it is quite possible that additional \he3 destruction mechanisms
(Charbonnel 1994, 1995, 1996; Hogan 1995; Wasserburg, Boothroyd, \&
Sackmann 1995) can lead to a consistent picture for the evolution of 
\he3 (Olive \etal 1997; Galli \etal 1997), one must argue that the new
process is not operative in all stars 
in order to avoid a contradiction between a few planetary nebulae
showing high \he3 abundances (from 
2 to $10 \times10^{-4}$, Balser \etal 1997; Balser, Rood, \& Bania 1999) 
 and \hii\ regions with small \he3 content
(about $2\times10^{-5}$, Balser \etal 1999). The \hii\ region
observations are in good agreement with the protosolar abundance value
${\rm \he3/H} = (1.5 \pm 0.2) \times 10^{-5}$ (Geiss \& Gloeckler 1998,
Gloeckler \& Geiss 1998). 
The problem of \he3 therefore seemed to be one  of stellar and/or Galactic in
nature. For this reason, it was deemed to be a poor cosmological tracer.

Recently, the conclusion that \he3 is not significantly produced 
in stars has been corroborated by a wealth of observations of 
\c13 anomalies in low mass RGB stars and in some planetary nebulae. 
The RGB $\c12/\c13$ anomalies point out the existence of extra mixing process
in low mass stars (Charbonnel \& do Nascimento 1998; Sackmann \&
Boothroyd 1999a, 1999b and references therein). In this context, 
the mechanism responsible for the
low $\c12/\c13$ observed in most of the RGB stars should lead to
destruction of \he3 in the external layers. The very large fraction of stars
experiencing this phenomenon (of order 90 \%)
seems to prevent the overproduction of \he3 in the course of Galactic evolution.
In addition, the \c12/\c13 ratio has been
recently observed in planetary nebulae.
While Palla et al.\ \cite{pgmst} find high \c12/\c13
inconsistent with mixing in one object,
Palla et al.\ \cite{pbstg} find 
that most of \c12/\c13 ratios in a sample of 14 planetary nebulae
are low, implying that extra mixing has occurred.
Balser, McMullin, \& Wilson (2002)
reach a similar conclusion in a study of 11 planetary nebulae.
These theoretical and observational 
results thus represent impressive progress in resolving the
\he3 problem,
and further 
studies are likely to clarify the situation more.
This rapid progress has reopened the question of whether
\he3 can be restored as a probe of the cosmic baryon density.
It will be shown below that
the present results are not sufficient by themselves
to constrain the Galactic evolution of \he3 and 
hence its primordial abundance.

Consequently, it appears necessary to take a careful look
at this problem, integrating the available data in a
Galactic evolutionary model to reevaluate quantitatively
the cosmological status of \he3. 
In section 2, we briefly review the most recent developments
in BBN that will be used in the present work.
In section 3, the evolution of \he3
is traced in the framework of a Galactic evolutionary model in order to
analyze the potential production/destruction of \he3 in stars.
We find that present observations and theory
cannot yet sufficiently 
constrain the  primordial abundance of \he3 at the
needed precision.
In section 4, we take advantage of recent CMB observations (and the
derived value of 
$\Omega_B h^2$) to draw some consequences on primordial \he3 and its evolution. 
Our summary and conclusions are given in section 5.

\section{Standard  Big Bang nucleosynthesis calculation of light elements abundances}
 
Standard BBN has been recently updated using new reaction rate
compilations (NACRE in particular, Angulo \etal 1999). Independent
results of this update are in good agreement (Vangioni-Flam \etal 2000,
Cyburt \etal 2001, Coc \etal 2002) and will be used here. From the
primordial abundance of \he3 deduced by Bania \etal (2002) \he3/H $= (1.1
\pm 0.2)\times 10^{-5}$, we derive through
the BBN calculations the corresponding baryon-to-photon ratio
$\eta_{10} = \eta/10^{-10}$, as well as the D, \he4, and
\li7 primordial abundances.  These values are given in Table 1.
The last column presents the stellar destruction
factor of \li7 required to bring the primordial abundance down to the
Spite plateau  (see e.g., Spite \etal 1996). Here we have taken the
Plateau value to be ${\rm Li/H}  = 1.2 \times 10^{-10}$ (Ryan \etal 2000).
In addition, we explore a more extended range in $\eta_{10}$  taking into
account the more conservative limits of Bania \etal (2002). The models we
consider cover the range $\eta_{10}$ = 2.2 to 6.5.

\begin{table}[t]
\caption{BBN calculations of light nuclei abundances as a function of
$\eta$. }
\footnotesize
\begin{center}
\begin{tabular}{|cc|ccccc|c|}
\hline
\hline
 Model & $\eta_{10}$ & $Y(\he4)^{a}$ & $10^{5}\times$D/H$^b$ &
$10^{5}\times$\he3/H$^b$ & 
  $10^{10}\times$\li7/H$^b$ & \li7 Depletion$^c$\\
\hline
1 & 6.5 & 0.249 & 2.5& 0.9 & 4.2 & 3.4\\
2 & 4.7 & 0.246 & 4.1 & 1.1 & 2.3 & 1.9\\
3 & 3.7 & 0.243 & 5.8 & 1.3 & 1.5 & 1.2\\
4 & 3.1 & 0.241 & 7.9 & 1.5 & 1.2 & 1.0\\
5 & 2.2 & 0.237 & 13.0 & 1.9 &1.3 & 1.1\\
\hline
\end{tabular}
\end{center}
\noindent
{\it a}: mass fraction \\
{\it b}: number with respect to H \\
{\it c}: the stellar destruction factor that would be needed
to bring Li down to the observed Spite plateau.

\end{table}

Each of the light elements can be individually
used to determine $\Omega_b h^2$ with varying accuracy.  
Table 1 illustrates the weak dependence of \he3 on $\eta$:
a factor of 2 variation in \he3 corresponds
to a factor of 3 variation in $\eta$.  
This already foreshadows the difficulty of using \he3 to
determine the baryon density.
\he4 displays an even weaker, logarithmic dependence on density, 
so that very accurate abundance
measurements are necessary to use \he4 as a baryometer.  At present,
systematic uncertainties make it
difficult to exclude values of Y as high as 0.25 
(Olive \& Skillman, 2001) thus allowing $\eta_{10} \le 7.0$.
With \li7 one can do much better. A plateau value of ${\rm \li7/H} = 1.2
\times 10^{-10}$ corresponds to $\eta_{10} =  2.4$ and 3.2.  However, the
slightly higher value of ${\rm \li7/H} = 1.9 \times 10^{-10}$ (consistent
within systematic uncertainties) yields $\eta_{10} = 1.7$ and 4.3.

Indeed, the question of the lithium depletion during the pre-main
sequence and main sequence phases of stellar evolution has been a long
standing problem. Historically, the range of possible depletion factors
has been steadily decreasing over the years. Some ten years ago, it
ranged from about 0.2 dex in standard non rotating stellar models to an
order of 1 dex in models with rotational mixing; now, the estimated
depletion factor lies between 0.2 and 0.4 dex (Pinsonneault \etal 1999,
2001). While there is no lack of processes which could lead to lithium
depletion in the envelope of stars: mixing induced by rotation,
microscopic diffusion and mass loss (Vauclair \& Charbonnel 1998; Theado
\& Vauclair 2001), however, none of these processes are free from
objection; more specifically, the understanding of the behavior of the
angular momentum vs radius and time and its transport is far from complete.
Independent of these theoretical arguments,  the observational data
indicate very little if any depletion of Li. First, there is the total 
lack of dispersion (beyond observational uncertainty) in the Lithium data
(Bonifacio \& Molaro 1997;  Ryan, Norris,
\& Beers 1999; Ryan \etal 2000), and second,  recent
\li6 observations also leave little room, if any, for \li6
depletion (Fields \& Olive 1999; Vangioni-Flam \etal 1999).
 
In contrast, the strong dependence of D/H on $\eta$ makes deuterium a
very good baryometer, provided we have an accurate determination of D/H.
The values of D/H measured in high redshift systems however cover
the range $\eta_{10} = 4.8$ for ${\rm D/H} = 4.  \times 10^{-5}$ to $\eta_{10}=8.5$ for
${\rm D/H} = 1.65\times 10^{-5}$ (Burles \& Tytler 1998a, 1998b, Tytler
2000, O'Meara \etal 2001, D'Odorico \etal 2001, Pettini \& Bowen 2001).
We note that if the dispersion in the D/H is real
(beyond observational error as the quoted error bars imply), then one
can question the extent that any of these determinations truly
represent the primordial D/H abundance and some amount of D/H must have
been destroyed, presumably by stellar processing.
Since ${\rm D/H}(t)$ is monotonically
decreasing in time, the primordial value should be greater than or
equal to the maximum of the  observed values (Fields \etal 2001).

\section{Stellar and Galactic evolution of \he3}

The aim of this work is to check the viability of \he3 as a baryometer. In this
context, we have run standard Galactic evolutionary  models which in essence
follow the evolution of the abundance of the elements in the ISM, and thus of
the \hii\ regions, starting from their estimate of primordial abundances
(Table 1). Five models are considered with \he3 primordial abundances
varying from 0.9 to 1.9 $\times 10^{-5}$
(corresponding $\eta_{10} = 6.5$ to 2.2). 
Relying on recent work concerning the behaviour of \he3 in stars, we
attribute only moderate production of \he3 in stars with masses
between 1 and 3 M$_\odot$ Indeed, the rare
observations of \he3 in planetary nebulae (Balser \etal 1997, 1999)
lead to \he3/H  $\approx 10^{-3}$. As pointed out by Olive \etal (1997)
and Galli \etal (1997) on the basis of Galactic
evolutionary models, only a limited fraction of low mass stars can be
\he3 producers. This conclusion has been recently corroborated by stellar
evolutionary models (Charbonnel \& do Nascimento 1998 and Sackmann \&
Boothroyd 1999) implying a fraction of about 10 \% for \he3 producers. In
this context, the \he3 yields are taken to be  1 to 5 $\times 10^{-4}$
M$_\odot$ in the  mass range considered above.  However, due to the
inherent uncertainties in the physics governing \he3 production, there is
obviously a certain amount of leeway on the yields which does not
impact our general conclusions as shown below. 

The evolution of \he3 and Li vs [O/H] and [Fe/H] respectively, is shown in
Figure 1 over four decades in metallicity, for the five models considered.\footnote{
Note that for simplicity only the primordial \li7 component has been
included, and no Galactic sources of \li6 or \li7.  
These are important for Li evolution
at high metallicities (e.g., Vangioni-Flam \etal 1999;
Ryan et al.\ 2000),
but do not affect the comparison with the Spite plateau in the Figure.
}
As one can see, the \he3 data alone cannot discriminate between the five models
chosen.
The scatter in the points (whether real or observational)
makes it difficult to infer a mean \he3 evolutionary trend.
In addition, the relatively narrow range in metallicity
reduces the ``leverage'' of the data to reveal any trend.
For both of these reasons, 
the \he3 data alone do not allow one to extract a narrow range in $\eta_{10}$. 
The lithium data are more discriminative, and in particular, one model (the
solid curve corresponding to \he3/H$_{p} = 0.9 \times 10^{-5}$)
appears to be problematic and would require a depletion factor greater
than 3. The flat behaviour of \he3 vs [O/H] in the halo phase is due to
the lack of production of \he3 in massive stars. The moderate increase in
the disc phase is due to the release of \he3 by low mass stars.
Figure 2 zooms in on the results corresponding to the observational range
for [O/H].  Note that the higher the primordial \he3 the flatter the
evolutionary curve in the disc, for the same given contribution of low
mass stars. Note that, for a given \he3 primordial value, if we had chosen
a lower \he3 yield,
still consistent with the planetary nebula data, the evolutionary curve would  also be 
flatter.

\begin{figure}
\begin{center}
\epsfig{file=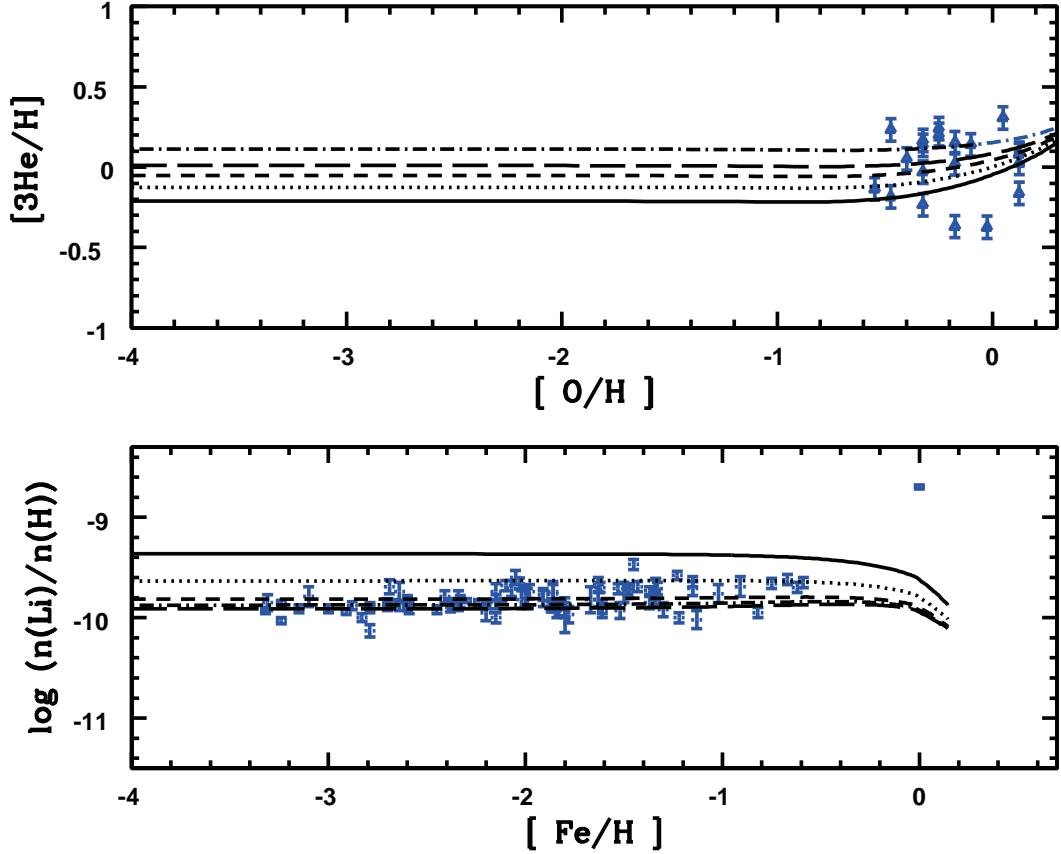,height=4.5in}
\end{center}
\caption{The evolution of \li7 and \he3 as a function of metallicity.
In the upper panel, we show the evolution of \he3 vs [O/H]. 
The different curves from bottom to top correspond to increasing \he3
primordial abundances from 0.9 to 1.9 $\times 10^{-5}$ (models 1 to 5 in
Table 1). The data points are from Bania \etal
(2002). In the lower panel, we show the evolution of the logarithmic
abundance of Li vs [Fe/H]. The different curves correspond to the models
given in Table 1 as per the \he3 abundances above. The data are from
Bonifacio \& Molaro (1997) and Ryan
\etal (1999).
\label{fig:E}
}
\end{figure}

\begin{figure}
\begin{center}
\epsfig{file=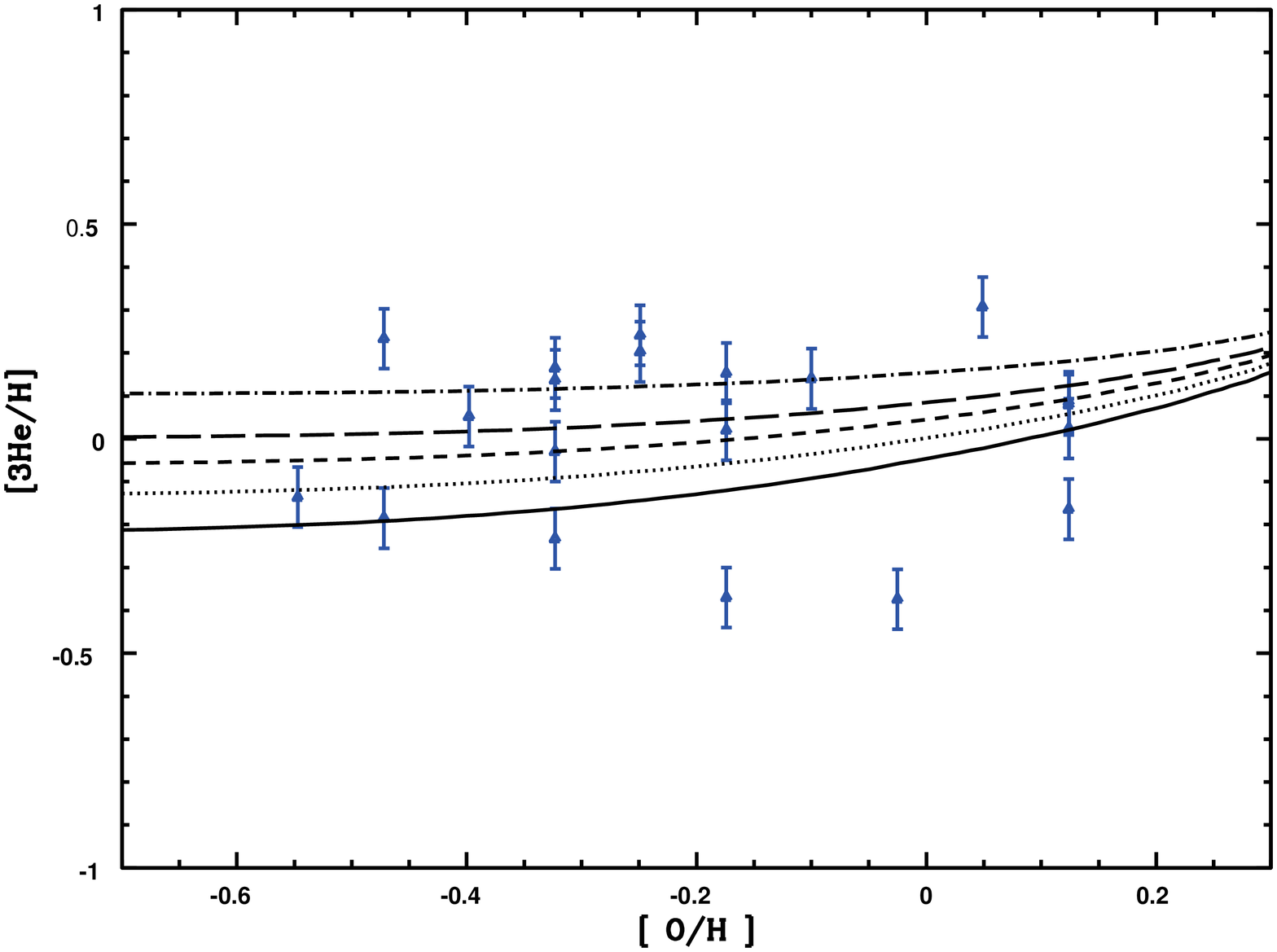,height=4.5in}
\end{center}
\caption{A blow-up of the evolution of \he3 in the
observed [O/H] range.
\label{fig:Z}
}
\end{figure}

 For illustrative purposes, we consider a model with total stellar 
destruction of \he3 as in case of deuterium and compare it with models
discussed above.  In Figure 3, we see that one
cannot discriminate between the two evolutionary assumptions. We
show the corresponding results of models 1, 3 and 5. The conclusion
that one can draw is simply that the current data are too scattered to
derive significant constraints on $\eta$, notwithstanding the
uncertainties in the stellar
\he3 production/destruction in low mass stars. The very existence of large
\he3 abundances in some planetary nebulae, however, favors a slight (but
rare) production of \he3, without implying any consequence on cosmology.

\begin{figure}
\begin{center}
\epsfig{file=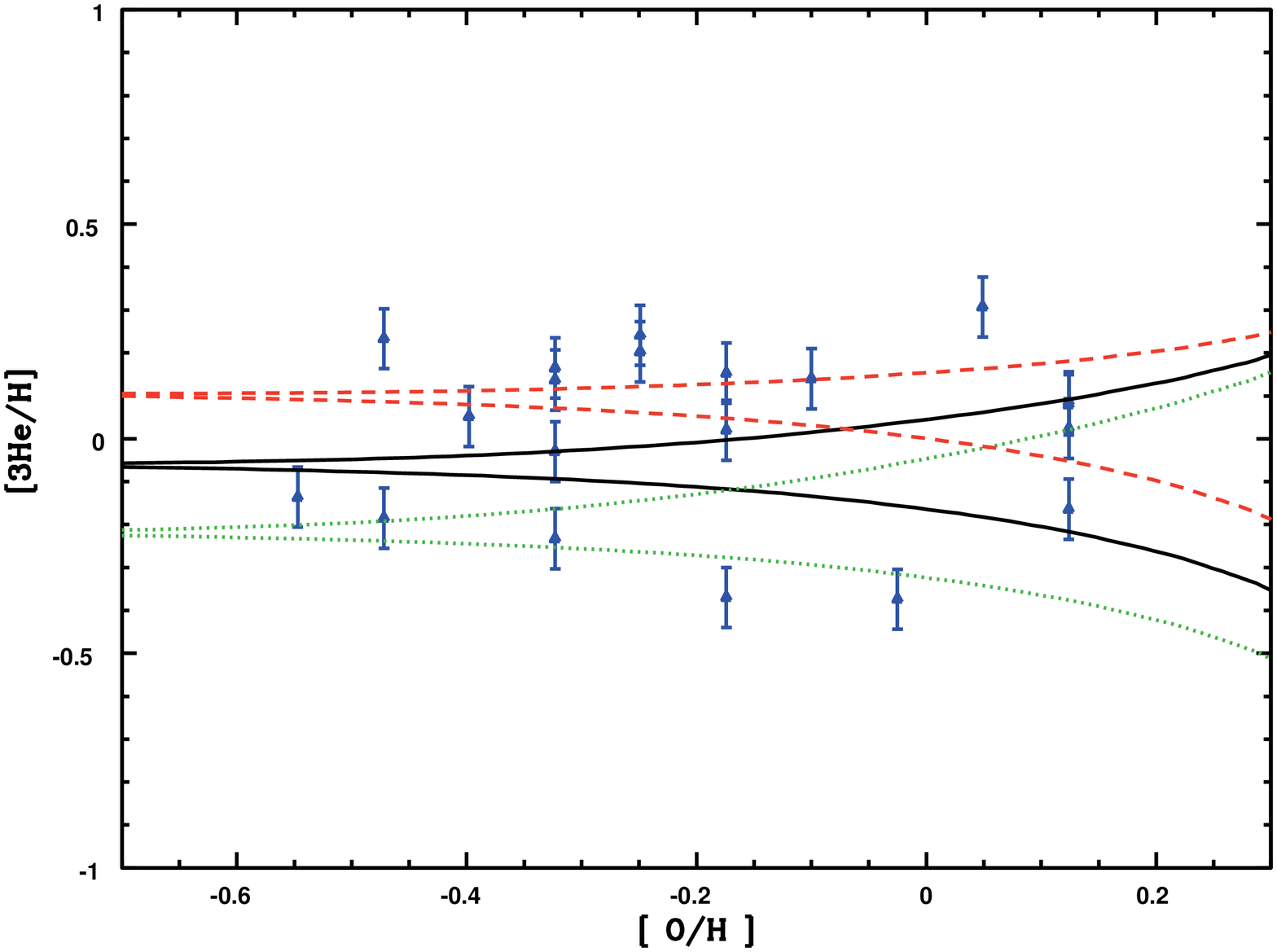,height=4.5in}
\end{center}
\caption{The evolution of \he3 vs [O/H] in the observed range.
The pairs of lines assume either a small production
of \he3 in the low mass range or stellar destruction only. The dotted,
solid and dashed curves correspond to models 1, 3 and 5 respectively. 
\label{fig:o}
}
\end{figure}

The analytic expressions for the chemical evolution of \he3
give some additional insight into the observed trends.
The mass fraction $X_3$ of \he3 evolves in a closed-box model as
$M_{\rm g} dX_3/dt = -X_3 E + E_3$.
Here $E = \int m_{\rm ej} \, \phi(m) \, \psi(t-\tau_m) \, dm$
and $E_3 = \int m_{\rm ej,3} \, \phi(m) \, \psi(t-\tau_m) \, dm$
are the usual mass and \he3 ejection functions (which includes deuterium-processing), 
and we have $\phi$ the IMF, $\psi$ the star formation rate,
and $\tau$ the stellar lifetime.
This expression can be recast as
\begin{equation}
\label{eq:dXdt}
dX_3/dt = - \Gamma_\star(X_3 - X_{3,{\rm eq}})
\end{equation}
with $\Gamma_\star(t) = E(t)/M_{\rm g}(t)$, and $X_{3,{\rm eq}}(t) = E_3(t)/E(t)$.
{}From eq.\ (\ref{eq:dXdt}) 
we see that 
\he3 is driven from its initial value towards
a steady state $dX_3/dt = 0$, in which $X_3$ approaches
$X_{3,{\rm eq}}$; 
this equilibrium value retains no memory of the primordial abundance.
If the primordial value
$X_{3,p}$ is higher (lower) than $X_{3,{\rm eq}}$, then 
the abundance will, over a timescale 
$\Gamma_\star^{-1} = M_{\rm g}/E \sim M_{\rm g}/\psi \sim 5$ Gyr,
be driven down (up) to the equilibrium
value.
The apparent flatness of \he3 versus O/H may suggest that such an equilibrium
has already been reached.
Or course, if the primordial value happens to be
$X_{3,p} \sim X_{3,{\rm eq}}$, then
the \he3 abundance will change very little over time.

Note that if one could firmly establish observationally that \he3 
definitely increases (decreases) with time, then
as Bania et al.\ \cite{brb02} argue, one could use the observed \he3
data to place an upper (lower) limit on the primordial value.
This is possible if one could determine the sign of 
$X_{3,{\rm eq}} - X_{3,p}$, perhaps by 
measurement of \he3 in an appropriate sample of planetary nebulae.
The limit one could derive on \he3 would still suffer from the
flatness of \he3 versus $\eta$, so that \he3 would not be
a precision baryometer.  On the other hand, precisely because of
this flatness, a firm limit on primordial \he3 would provide a useful consistency check on 
the baryon density arrived at by other means.
Indeed, with a determination of the baryon density, \he3
observations become very interesting for other reasons,
to which we now turn.

\section{Helium-3 with CMB Inputs:  Probing Stellar and Chemical Evolution}

CMB anisotropy data is now reaching the precision where it
can provide an accurate measure of the cosmic
baryon content.  This independent measure of  $\eta$
will test cosmology and will probably revolutionize BBN.
Given a CMB measurement of $\eta$, and presuming a BBN-CMB concordance,
one can turn the usual BBN problem around and predict the primordial
abundance of \he3. Because of the small \he3 variation with $\eta$, the
CMB data will provide a very precise prediction.

Indeed, an interesting prediction is already possible.
Recent results from DASI (Pryke \etal 2002) and CBI (Sievers \etal 2002)
indicate that $\Omega_B h^2 = 0.022^{+0.004}_{-0.003}$, while 
BOOMERanG-98 (Netterfield \etal 2002) gives 
$\Omega_B h^2 = 0.021^{+0.004}_{-0.003}$. These determinations are
somewhat lower than value found by MAXIMA-1 (Abroe \etal 2002)
 $\Omega_B h^2 = 0.026^{+0.010}_{-0.006}$ and VSA (Rubi\~no-Martin \etal
2002) $\Omega_B h^2 = 0.029 \pm 0.009$. Taking a CMB value
of $\Omega_B h^2 = 0.022 \pm 0.003$, or
$\eta_{\rm 10,cmb} = 6.0 \pm 0.8$ (1 $\sigma$), we would predict a 
primordial \he3 value of 
\he3/H in the range 0.87 to 1.04  $\times 10^{-5}$ with a central value
of 0.94 $\times 10^{-5}$, already a prediction good to  about 10 \%.
With MAP data, the accuracy of $\eta_{\rm cmb}$ should
be $10\%$ or better, which will give a \he3 prediction to
$6 \%$ or better.

This primordial \he3 abundance, while only one number,
provides a powerful probe for both chemical evolution and stellar evolution.
The primordial value provides a starting point
from which observed \he3 has evolved with metallicity.
Using chemical evolution to model the mean \he3-O trend,
as we have done here, one
can now probe stellar evolution.
Anchored at the initial \he3 value, 
different stellar yield assumptions will lead to strongly different 
present values, allowing one a new means of constraining 
possible stellar mixing models.

In addition, with a known primordial \he3,
one can then compare with individual observations in \hii\ regions, 
and determine whether \he3 has increased or decreased in that particular system.
Either way, this makes a statement about the interplay of stellar and \hii\ region evolution, 
which one can model to understand
the scatter of individual regions from the mean \he3-metallicity trend.
Such an analysis might be reminiscent of those proposed to explain
scatter in \he4-metal trends in \hii\ regions 
(e.g., Garnett \cite{don}; Pilyugin \cite{pil}).

\section{Conclusions}
 
Considerable observational and theoretical progress has recently been
made in understanding the stellar and Galactic evolution of \he3.
These studies have gone far to address the apparent dichotomy between 
the high \he3 values measured in some planetary nebulae, and 
the much lower values seen \hii\ regions.
Detailed stellar evolution models, and
observations of carbon isotopes in red giants and planetary nebulae,
strongly suggest that an extra mixing process
is responsible for the destruction of \he3 in
90\% of low mass stars; thus the apparent observational 
inconsistency is removed. 
A key new contribution to this emerging picture
are the Bania et al.\ (2002) measurements \he3 over a range
of metallicity; the observed lack of strong \he3 evolution with
oxygen confirms that 
the production of \he3 by low mass stars is very limited.

This progress in stellar physics notwithstanding, 
the remaining observational and theoretical uncertainties
surrounding \he3 evolution leave this isotope
poorly-suited to be
the precision baryometer found in \li7 and especially D.
On the other hand, \he3 nevertheless remains extremely interesting as an astrophysical probe.
More data will increase its value.  For example,
an unbiased survey of \he3 in planetary nebulae will shed
light on whether \he3 is increasing or decreasing from
its primordial value. Moreover, the
CMB results on $\eta$ allow one to predict primordial abundances quite
accurately for \he3 and the other light isotopes.  
This will cast the \hii\ region \he3 data in a new light, allowing these data to constrain both
stellar and chemical evolution.  
Thus, we foresee a promising future for
\he3, and urge that the longstanding \he3 observational program 
continue with renewed vigor.

\acknowledgments

We warmly thank Olivier Dor\'e 
and Anuj Sarma for helpful discussions.
This work was supported in part by DOE grants
DE-FG02-94ER-40823 at the University of Minnesota,
by NSF grant AST 00-92939 at the University of Illinois, 
and by PICS 1076 CNRS
France/USA.

\end{document}